\documentclass[11pt,fleqn]{article}

\usepackage{latexsym}

\usepackage{amsmath}
\usepackage{amsthm}
\usepackage{amssymb}
\usepackage{amsfonts}

\newcommand{\be}{\begin{equation} } 
\newcommand{\ee}{\end{equation} \par \noindent}

\newcommand{\rf}[1]{(\ref{#1})}

\newcommand{\scal}[2]{\mbox{$\langle #1 \! \mid #2 \rangle $}} 

\newcommand{\ba}{\begin{array}}
\newcommand{\ea}{\end{array}}
\newcommand{\const}{{\rm const}}

\newcommand{\e}{{\bf e}}
\newcommand{\C}{{\mathbb C}}
\newcommand{\R}{{\mathbb R}}
\newcommand{\E}{{\mathbb E}}

\newcommand{\m}{\left( \ba{c}}
\newcommand{\ema}{\ea \right)}
\newcommand{\mm}{\left( \ba{cc}}

\newtheorem{prop}{Proposition}

\newtheorem{rem}{Remark}

\newenvironment{Proof}{\par \vspace{2ex} \par
\noindent \small {\it Proof:}}{\hfill $\Box$ 
\vspace{2ex} \par }

\begin{document}

\title{\bf 
The Darboux-B\"acklund transformation for the
static 2-dimensional continuum Heisenberg chain}

\author{{\bf
Jan L.\ Cie\'sli\'nski\thanks{
E-mail: \tt janek\,@\,alpha.uwb.edu.pl}}
\\ {\bf Joanna Czarnecka}
\\ {\footnotesize Uniwersytet w Bia\l ymstoku, Instytut Fizyki Teoretycznej}
\\ {\footnotesize ul.\ Lipowa 41, 15-424  Bia\l ystok, Poland} 
}

\maketitle

\begin{abstract} 
We construct the Darboux-B\"acklund transformation for 
the sigma model describing static configurations of 
the 2-dimensional classical continuum Heisenberg chain. The transformation is characterized by a non-trivial normalization matrix depending on the background solution. In order to obtain the transformation we use a new, more general, spectral problem. 
\end{abstract}

\par \vspace{0.5cm} \par
\par \vspace{0.5cm} \par
\par \vspace{0.5cm} \par

\section{Introduction}

In this paper we consider 2-dimensional Euclidean $O(3)$ $\sigma$-model 
\be  \label{Nmodel}
     n,_{xx} + n,_{yy} + (n,_x^2 + n,_y^2) n = 0 \ , \qquad n^2 = 1 \ ,
\ee
where $n \in \E^3$. 
This equation appears in the classical field theory 
and solid state physics. 
Considering $2+1$-dimensional $S^2$ $\sigma$-model  \cite{LPZ}
\be
 \partial^\mu \partial_\mu \phi + (\partial^\mu \phi \cdot \partial_\mu \phi) \phi \ ,
\qquad \phi \cdot \phi = 0 \ ,
\ee
and $2+1$-dimensional continuum classical Heisenberg ferromagnet equation 
\be
     {\vec S},_t = {\vec S} \times ({\vec S},_{xx} + {\vec S},_{yy}) \ , \qquad {\vec S}^2 = 1 \ ,
\ee
we see that their static solutions satisfy \rf{Nmodel}. 

This $\sigma$-model plays important role also in differential geometry. The normal vector $n$ to surfaces of constant mean curvature endowed with conformal coordinates satisfies \rf{Nmodel} \cite{Bob,CS}.

There are many interesting papers on the interpretation of the sigma model \rf{Nmodel} and on the construction of special solutions \cite{BePo,Bob,BGM,DS,SW}. In this paper we present a large family of gauge-equivalent spectral problems associated with the $\sigma$-model \rf{Nmodel} and construct the Darboux-B\"acklund transformation for this general spectral problem.

\section{The spectral problem}
\label{spectral}

We consider the spectral problem of the form
\be  \ba{l}  \label{problin-genH}
\displaystyle
   \Psi,_x = U  \Psi \equiv   \left( A \zeta -  \frac{A^\dagger}{\zeta} + R \right) \Psi \ , \\[3ex] 
\displaystyle
\Psi,_y =  V \Psi \equiv  \left( B \zeta - \frac{B^\dagger}{\zeta} + S \right) \Psi \ ,
\ea \ee
(where $U,V,\Psi$ are $2\times 2$ matrices)  
uniquely characterized by the following properties:
\begin{itemize}
\item[(A)]\quad  $U, V$  are rational in $\zeta$ with simple poles at $\zeta = 0$ and $\zeta = \infty$,
\item[(B)]\quad  $ ( U ( 1 / \zeta) )^\dagger = - U (\bar \zeta) $ \ , \quad $ ( V ( 1/ \zeta) )^\dagger = 
 - V ( \bar \zeta) $
\item[(C)]\quad $A^2 = B^2 = 0$ \ ,  
\item[(D)]\quad  $B = i A$ \ .
\end{itemize}
The constraint (B) implies $R^\dagger = - R$, $S^\dagger = - S$, i.e., $R$, $S$ are $u(2)$-valued.
The compatibility conditions (the coefficient by $\lambda^2$) imply that $A$ and $B$ are parallel, i.e.,
\be  \label{ABW}
   A = a W \ ,  \qquad B = b W \ , 
\ee
where $a, b \in \C$. Without loss of the generality we can assume $a \in \R$ and 
\be  \label{asaW}
   a > 0 \ , \qquad  \scal{W}{W^\dagger} = - 2 \ ,
\ee
where the scalar product on the space of $2\times 2$ matrices is defined by $\scal{X}{Y} = - 2 {\rm Tr} (X Y)$.  
The coefficient $2$ assures that the basis $\e_k \equiv - i \sigma_k/2$ is orthonormal. We use the standard representation of Pauli matrices, i.e., 
\be  \label{ePauli}
  \e_1 = \frac{1}{2} \mm 0 & - i \\ - i & 0  \ema  , \quad 
\e_2 =  \frac{1}{2} \mm    0 & - 1 \\ 1 & 0 \ema  , \quad
\e_3 =  \frac{1}{2} \mm    - i & 0 \\ 0 & i \ema  .
\ee
The assumptions \rf{asaW} make the choice of $W$ in equations \rf{ABW} unique. Indeed, $\scal{A}{A^\dagger} = |a|^2 \scal{W}{W^\dagger} = - 2 |a|^2$. 
Therefore $|a|^2 = - \scal{A}{A^\dagger}/2 = {\rm Tr} (A A^\dagger)$ and, finally,
\be  \label{aW}
   a = \sqrt{{\rm Tr} (A A^\dagger)} \ , \qquad W = A / a \ .
\ee
The constraint (D) reduces to $b = i a$ and is necessary to obtain the standard form of the Laplace operator (geometrically it means that we choose conformal coordinates on the corresponding constant mean curvature surface).

It is convenient to define the following frame
\be \label{ortho}
     E_1 = \frac{W + W^\dagger}{2 i} \ , \qquad E_2 = \frac{W^\dagger - W}{2} \ , \qquad E_3 = [ E_1, E_2] \ .
\ee
Note that $\scal{E_k}{E_j} = \delta_{kj}$, $E_k^\dagger = - E_k$ and ${\rm Tr} E_k = 0$ for $k,j=1,2,3$. 
Thus this is an orthonormal basis in $su(2)$.
Any orthonormal basis in $su(2)$ can be parameterized by a wector $W$, satisfying $W^2 = 0$ and ${\rm Tr} (W W^\dagger) = 1$, according to the formulas \rf{ortho}.

The kinematics of the frame $E_1, E_2, E_3$ can be 
expressed in terms of six functions 
($\alpha_1, \beta_1, \gamma_1, \alpha_2, \beta_2, \gamma_2$):
\be \ba{l}  \label{kinem}
\displaystyle \frac{\partial }{\partial x} \m E_1 \\ E_2 \\ E_3 \ema = 
\left( \ba{ccc} 
0 & \alpha_1 & \beta_1 \\ - \alpha_1 & 0 & \gamma_1 \\ 
- \beta_1 & - \gamma_1 & 0 
\ea \right) \m E_1 \\ E_2 \\ E_3 \ema  \\[5ex]
\displaystyle \frac{\partial }{\partial y}  \m E_1 \\ E_2 \\ E_3 \ema  = 
\left( \ba{ccc} 
0 & \alpha_2 & \beta_2 \\ - \alpha_2 & 0 & \gamma_2 \\ 
- \beta_2 & - \gamma_2 & 0 
\ea \right) \m E_1 \\ E_2 \\ E_3 \ema 
\ea \ee 
satisfying the compatibility conditions:
\be  \ba{l}  \label{cc0}
\alpha_1,_y - \alpha_2,_x + \beta_2 \gamma_1 - \beta_1 \gamma_2 = 0 \ , \\[2ex]
\beta_1,_y - \beta_2,_x + \alpha_1 \gamma_2 - \alpha_2 \gamma_1 = 0 \ , \\[2ex]
\gamma_1,_y - \gamma_2,_x + \alpha_2 \beta_1 - \alpha_1 \beta_2 = 0 \ . 
\ea \ee
Denoting $\zeta = \exp (- i \kappa)$ and expressing $U, V$ in terms of $E_k$, we rewrite the spectral problem \rf{problin-genH} as follows:
\be \ba{l}  \label{UV}
U = a E_2 \cos\kappa - a E_1 \sin\kappa   + R \ , \\[2ex]
V = a E_2 \sin\kappa + a E_1 \cos\kappa + S \ , 
\ea \ee
The compatibility conditions for the spectral problem 
\rf{UV} read 
\be \ba{l} \label{cc}
(a E_2),_y - (a E_1),_x + a [E_2, S] - a [E_1, R] = 0 \ , \\[2ex]
(a E_2),_x + (a E_1),_y + a [E_1, S] + a [E_2, R] = 0 \ , \\[2ex]
R,_y - S,_x + [R, S] = a^2 E_3 \ .
\ea \ee
The $u(2)$-valued functions $R$ and $S$ are linear combinations of $iI, E_1, E_2, E_3$ (with real coefficients), i.e., $S = i s_0 + s_1 E_1 + s_2 E_2 + s_3 E_3$ and
$R = i r_0 + r_1 E_1 + r_2 E_2 + r_3 E_3$. Thus the system \rf{cc} can be written in a more explicit form:
\be \label{ccc}
\ba{l}
r_1 - \gamma_1 = \beta_2 + s_2 \ , \qquad
r_2 + \beta_1 = \gamma_2 - s_1 \ , \\[2ex]
a,_x + a \alpha_2 - a s_3 = 0 \ , \quad
a,_y - a \alpha_1 + a r_3 = 0 \ , \quad r_0,_y = s_0,_x \ ,\\[2ex]
r_1,_y - s_1,_x + r_2 s_3 - r_3 s_2 - \alpha_2 r_2 - \beta_2 r_3 + \alpha_1 s_2 + \beta_1 s_3 = 0 \ , \\[2ex]
r_2,_y - s_2,_x + r_3 s_1 - r_1 s_3 + \alpha_2 r_1 - \gamma_2 r_3 - \alpha_1 s_1 + \gamma_1 s_3 = 0 \ , \\[2ex]
r_3,_y - s_3,_x + r_1 s_2 - r_2 s_1 + \beta_2 r_1 + \gamma_2 r_2 - \beta_1 s_1 - \gamma_1 s_2 = 4 a^2 \ .
\ea \ee 

\begin{prop} \label{main}
Let $\Psi$ satisfies \rf{problin-genH}, (A)-(D), and \rf{ortho} holds. Then
\be
 n = \Psi^{-1} E_3 \Psi 
\ee
satisfies Eq.\ \rf{Nmodel}, i.e., 
$  n,_{xx} + n,_{yy} = f (x,y)\  n$, 
where $f$ is a real function. 
\end{prop}

\begin{Proof}
We compute $n,_{xx} + n,_{yy}$ and obtain, as a consequence, a linear combination of $\Psi^{-1} E_k \Psi$ ($k=1,2,3$). 
It is enough to show that the result is proportional to $\Psi^{-1} E_3 \Psi$, i.e., that the coefficients by $\Psi^{-1} E_1 \Psi$ and 
$\Psi^{-1} E_2 \Psi$ vanish. The coefficient by $\Psi^{-1} E_2 \Psi$ 
is given by
$$  \ba{l}
(r_1 - \gamma_1),_x - (\gamma_2 - s_1),_y + (\beta_1 + r_2) ( r_3 - \alpha_1) + (\beta_2 + s_2) (s_3 - \alpha_2) \\[2ex]
+ ( a,_y - a \alpha_1 + a r_3 ) \cos\kappa +  ( a,_x + a \alpha_2 - a s_3 ) \sin\kappa  \ .
\ea $$
To show that this expression vanish we use the first two equations of the system \rf{ccc}, then we eliminate all derivatives using appropriate equations of \rf{cc0} and \rf{ccc}.  Using once more (if necessary) 
the first equation of \rf{ccc} we see that the obtained result is zero. The coefficient by $\Psi^{-1} E_1 \Psi$ 
$$ \ba{l}
- (r_2 + \beta_1),_x  - (\beta_2 + s_2),_y + (s_1 - \gamma_2)(s_3 - \alpha_2) + (r_3 - \alpha_1)(r_1 - \gamma_1) \\[2ex]
- (a,_x + a \alpha_2 - a s_3) \cos\kappa - (a,_y + a r_3 - a \alpha_1) \sin\kappa 
\ea $$
vanishes as well what can be shown in exactly the same way.  
\end{Proof}

\begin{rem}    \label{isom}
The coefficients of the matrix $n$ with respect to the basis \rf{ePauli}, $n = n_1 \e_1 + n_2 \e_2 + n_3 \e_3$, identify this matrix with an $\E^3$ vector $(n_1, n_2, n_3)$. In other words, we use the isomorphism between $su(2)$ and $\E^3$.
\end{rem}

\begin{rem}
If one more constraint, namely ${\rm Tr} U = {\rm Tr} V = 0$, $\det \Psi = 1$, is imposed on the linear problem \rf{problin-genH}, then the Sym-Tafel formula $F = \Psi^{-1} \Psi,_\kappa$ yields surfaces of constant mean curvature, compare \cite{Ci-nos,DS}. 
\end{rem}

\section{Gauge transformations}

The spectral problem \rf{problin-genH} is invariant with respect to gauge transformations of the form $\hat \Psi = G \Psi$, where $G$ is any $\zeta$-independent $U(2)$-valued  matrix ($G^{-1} = G^\dagger$). 

\begin{prop}  \label{G}
If $\Psi$ satisfies \rf{problin-genH}, (A)-(D), and $\hat \Psi = G \Psi$, where $G^{-1} = G^\dagger$, then $\hat \Psi$  satisfies \rf{problin-genH}, (A)-(D) as well. Moreover
\be
\hat n \equiv  \hat \Psi^{-1} {\hat E}_3 \hat \Psi = n \ .
\ee
\end{prop}
\begin{Proof} \quad ${\hat \Psi},_x = {\hat U} {\hat \Psi}$, 
${\hat \Psi},_y = {\hat V} {\hat \Psi}$, where
\[  \ba{l} \displaystyle
 \hat U =  G A G^{-1} \zeta -  \frac{ G A^\dagger G^{-1} }{\zeta} + R +  G,_x G^{-1} \equiv  \hat A  \zeta -  \frac{ {\hat A}^\dagger }{\zeta} + \hat R  \ , \\[4ex] \displaystyle
\hat V = G B G^{-1} \zeta - \frac{ G B^\dagger G^{-1} }{\zeta} + S + G,_y G^{-1}   \equiv  \hat B  \zeta - \frac{ {\hat B}^\dagger }{\zeta} + \hat S    \ ,
\ea \]
where $(G A G^{-1})^\dagger = G A^\dagger G^{-1}$ because $G^\dagger = G^{-1}$. Obviously, ${\hat A}^2 = {\hat B}^2 = 0$, ${\hat R}^\dagger = - \hat R$, 
${\hat S}^\dagger = - \hat S$, etc. Thus the matrices $\hat U, \hat V$ satisfy all conditions (A)-(D). Then
$ {\rm Tr} ({\hat A} {\hat A}^\dagger) = {\rm Tr} ( G A G^{-1} G A^\dagger G^{-1}) = {\rm Tr} (A A^\dagger)$. Hence, taking into account \rf{aW} and \rf{ortho}, we get ${\hat a} = a$, ${\hat W} = G W G^{-1}$ and ${\hat E}_k = G E_k G^{-1}$ ($k=1,2,3$). Finally, 
${\hat n} = \Psi^{-1} G^{-1} G E_3 G^{-1} G \Psi = n$. 
\end{Proof}

\begin{prop} There exists a matrix $G = G (x,y) \in U(2)$ transforming the spectral problem \rf{problin-genH}, (A)-(D) into
\be  \ba{l}  \label{problinH}
\displaystyle
   {\hat \Psi},_x =  \left(  a \e_+ \zeta -  \frac{ a \e_-}{\zeta} + \hat R \right) \hat \Psi \ , \\[4ex] 
\displaystyle
{\hat \Psi},_y =   \left( i  a \e_+ \zeta + \frac{i a \e_-}{\zeta} + \hat S \right) \hat  \Psi \ ,
\ea \ee
where $a$ is given by \rf{aW} and 
\be  \label{epm}
\e_+ = \mm 0 & 1 \\ 0 & 0 \ema \ , \qquad \e_- = \mm 0 & 0 \\ 1 & 0 \ema \ .
\ee
\end{prop} 
\begin{Proof} 
Any two orthonormal bases in $\E^3$ are related by an orthogonal transformation, which in turn can be represented by a unitary matrix (the spinor representation). In particular, the basis $E_1, E_2, E_3$ from Section~\ref{spectral} can be obtained from any constant orthonormal basis $\e_1, \e_2, \e_3$ by the transformation of the form  $E_k = G^{-1} \e_k G$, $G \in U(2)$ (or even $G \in SU(2)$, if both bases have the same orientation). Applying the gauge transformation ${\hat \Psi} = G \Psi$ to the spectral problem \rf{problin-genH} we obtain \rf{problinH}, where 
$\e_+, \e_-$ are constant matrices such that $\e_- = \e_+^\dagger$, $\e_+^2 = \e_-^2 = 0$, $\scal{\e_+}{\e_-} = -2$. If $\e_k$ are given by \rf{ePauli}, then $\e_\pm$ are given by \rf{epm}. 
\end{Proof}

\begin{rem}  \label{solNe}
If $\hat \Psi$ solves \rf{problinH}, then $n = {\hat \Psi}^{-1} \e_3 \hat \Psi$ satisfies \rf{Nmodel}. 
\end{rem}

\begin{rem}
The spectral problem \rf{problinH} or its equivalents are usually applied in the spectral approach to constant mean curvature surfaces, see \cite{Bob,Ci-nos,DS}.  
\end{rem}

\section{The Darboux-B\"acklund transformation}

Our aim is to construct the transformation $\tilde \Psi = D \Psi$ (where $D$ depends on $x,y$ and $\zeta$) 
in such a way that ${\tilde U} = D,_x D^{-1} + D U D^{-1}$ and 
${\tilde V} = D,_y D^{-1} + D V D^{-1}$ 
have the same form as $U, V$
(compare \cite{Ci-dbt}). In other words, the properties (A), (B), (C), (D) of Section~\ref{spectral}
should be preserved by the transformation.
We confine ourselves to the simplest case
\be
    D = {\cal N} \left( I + \frac{\zeta_1 - \mu_1}{\zeta - \zeta_1} P \right) 
\ee
where the matrices ${\cal N}$ and $P$ do not depend on $\zeta$, $P^2 = P$, and $\zeta_1$, $\mu_1$ are 
complex parameters ($\zeta_1 \neq \mu_1$).

The property (A) implies, by virtue of a well known result of Zakharov and Shabat \cite{ZMNP}, 
\be
    {\rm ker} P \ni \Psi (\zeta_1) {\vec b} \ , \qquad {\rm im} P \ni \Psi (\mu_1) {\vec c} \  ,
\ee
where ${\vec b}, {\vec c} \in \C^2$ are constant vectors 
and $\zeta_1 , \mu_1 \in \C$ are constant as well.

One can easily check that the property (B) is preserved if $D^{-1} (\bar \zeta) = D^\dagger (1/\zeta)$ 
which yields, after straightforward computations, 
\be
   P^\dagger = P \ , \qquad  {\bar \mu}_1 = \frac{1}{\zeta_1} \ , \qquad {\cal N} {\cal N}^\dagger = 1 + (|\zeta_1|^2 - 1) P \ .
\ee
Therefore the condition $\mu_1 \neq \zeta_1$ is equivalent to $|\zeta_1| \neq 1$.  Moreover, $P^\dagger = P$ implies ${\vec c}_1 \perp {\vec b}_1$. 
$P$ is explicitly expressed by the matrix $\Psi (1/\bar\zeta_1)$:
\be  \label{pepsi}
   P = \frac{1}{1 + |\xi|^2} \mm |\xi|^2 & \xi \\ \bar \xi & 1 \ema \ , 
\ee 
where  $\xi = u_1/u_2$ and $(u_1, u_2)^T = \Psi (1/\bar \zeta_1) \vec c$.
One can check that the equation ${\cal N} {\cal N}^\dagger = 1 + (|\zeta_1|^2 - 1) P $ is satisfied by
\be  \label{normalization}
  {\cal N} = {\cal N}_0 (I + (\zeta_1 e^{i \sigma } - 1 ) P ) 
\ee
where ${\cal N}_0$ is a unitary matrix (${\cal N}_0^{-1} = {\cal N}_0^{\dagger}$) and $\sigma$ is a real constant. 

Considering the spectral problem \rf{problinH} we have to take into account one more constraint: $W = \e_+ $ is a fixed constant matrix,  given for instance by \rf{epm}. In this case
\be  \label{additional}
\tilde a \e_+ = a {\cal N} \e_+ {\cal N}^{-1} 
\ee
and from \rf{additional} we can compute ${\cal N}_0$. 

In the following we focus on the more general spectral problem \rf{problin-genH} and the matrix ${\cal N}_0$ can be arbitrary. Actually, the matrix ${\cal N}_0$ is not important as far as the transformation of $n$ is concerned (compare Proposition~\ref{G}). Without loss of the generality we will assume ${\cal N}_0 = I$.
Finally we arrive at following formula for the Darboux matrix:
\be
   D = \left(I + (\zeta_1 e^{i \sigma } - 1 ) P \right) \left( I + \frac{\zeta_1 - {\bar \zeta}_1^{-1}}{\zeta - \zeta_1} P \right) = I + ( e^{2 i \beta} - 1 ) P \ ,
\ee
where 
\be
e^{2 i \beta} := \frac{ (\zeta_1 - \zeta |\zeta_1|^2 ) e^{ i \sigma_1} }{ |\zeta_1|^2 - \zeta {\bar \zeta}_1 } \ . 
\ee
Note that $\beta$ is real (because $\bar \zeta = \zeta^{-1}$) and $\beta$ does not depend on $x, y$.

One can always parameterize the Hermitean projector $P$ by a unit vector ${\vec p } = (p_1,p_2,p_3)$:
\be
P = \frac{1}{2} (I + {\mathbf p} ) \ , \qquad {\mathbf p} :=\sum_{k=1}^3 p_k \sigma_k \ , \qquad    p_1^2 + p_2^2 + p_3^2 = 1 \ .
\ee
The function $\xi$ appearing in \rf{pepsi} is a stereographic projection of ${\vec p}$:
\be  \label{peksi}
   \xi = \frac{p_1 - i p_2}{1 - p_3} \ , \quad  p_1 = \frac{ 2 {\rm Re} \xi }{1 + |\xi|^2} \ , \quad p_2 = \frac{ - 2 {\rm Im} \xi }{1 + |\xi|^2} \ , \quad p_3 =  \frac{ |\xi|^2 - 1}{|\xi|^2 + 1} \ .
\ee

The spectral problem \rf{problinH} can be considered as a particular case of \rf{problin-genH} and any solution $\Psi$ of \rf{problinH} satisfies \rf{problin-genH} as well, compare Remark~\ref{solNe}. Therefore we can take as a background solution 
\be \label{ne3}
{\mathbf n} = \Psi^{-1} \e_3 \Psi \ , 
\ee
where $\Psi$ is a solution of \rf{problinH}. According to Remark~\ref{isom} we associate with ${\mathbf n}$ a unit vector ${\vec n} := (n_1, n_2, n_3)$ defined by:
\be
 {\mathbf n} = \sum_{k=1}^3 n_k \e_k \ . 
\ee
The Darboux-B\"acklund transformation of ${\mathbf n}$ yields:
\be  \label{tilden}
{\mathbf {\tilde n} } = \Psi^{-1} D^{-1} \e_3 D \Psi \ .
\ee
The obtained expression can be computed as follows
\[
D^{-1} \e_3 D = \frac{1}{2 i} (\cos\beta - i {\mathbf p } \sin\beta ) \sigma_3 (\cos\beta + i {\mathbf p } \sin\beta  ) \ ,
\]
and simplified in a straightforward way:
\be  \label{DeD} D^{-1} \e_3 D = 
\frac{1}{2 i} \left( \sigma_3 \cos 2\beta + 2 p_3 {\mathbf p} \sin^2 \beta + (p_2 \sigma_1 - p_1 \sigma_2) \sin 2 \beta \right) \ .
\ee

\section{Special solutions}

We will compute explicitly the action of the Darboux-B\"acklund transformation on a simple   background. 
The simplest seed solution can be obtained  from the requirement 
$U = \const, V =\const$ and $\Psi$ satisfies \rf{problinH}. Then $E_k = \e_k$ are constant 
(i.e., $\alpha_k = \beta_k = \gamma_k = 0$) and $a = a_0 = \const$. Thus the system 
\rf{ccc} reduces to
\be
r_1 = s_2 \ , \quad r_2 = - s_1 \ , \quad s_3 = r_3 = 0 \ , \quad
r_1 s_2 - r_2 s_1 = a_0^2 \ ,
\ee
and can be easily solved:
\be
  s_1 = - r_2 = a_0 \cos\delta_0 \ , \quad r_1 = s_2 = a_0 \sin\delta_0  \ ,
\ee
where $\delta_0$ is an arbitrary real parameter. Therefore,
\be  \ba{l}
  U = a_0 \e_2 ( \cos\kappa - \cos\delta_0) - a_0 \e_1 (\sin\kappa - \sin\delta_0)    \ , \\[2ex]
V = a_0 \e_1 ( \cos\kappa + \cos\delta_0) + a_0 \e_2 ( \sin\kappa + \sin\delta_0)  \ , 
\ea \ee
and, finally
\be  \ba{l}
U = 2 a_0 \sin\delta_- ( \e_1 \cos\delta_+ + \e_2 \sin\delta_+ ) \ , \\[2ex]
V = 2 a_0 \cos\delta_- (\e_1 \cos\delta_+ + \e_2 \sin\delta_+ ) \ ,
\ea \ee
where $\delta_\pm : = \frac{1}{2} (\delta_0 \pm \kappa) $. Without loss of the generality we put $\delta_0 = 0$ (more general choice corresponds to symmetries of the sigma model \rf{Nmodel} like rotation in 
the space of parameters $x,y$ and the $O(3)$ symmetry). Then
\be
  - 2 \e_1 \cos\delta_+ - 2 \e_2 \sin\delta_+ = i \mm 0 & e^{- i\kappa/2} \\ e^{i\kappa/2} & 0 \ema =: E \ .
\ee
Note that $E^2 = -1$.
Therefore, if $U, V$ are constant, then the solution of the linear problem \rf{problin-genH} is simply given by 
\be
\Psi = \exp (x U + y V) C_0 = \exp (\theta E) C_0 = (\cos\theta + E\sin\theta ) C_0 \ ,
\ee
where $C_0$ is a constant unitary matrix and 
\be  \label{teta}
\theta = \theta(x,y,\zeta) = a_0 x \sin \frac{\kappa}{2} - a_0 y \cos\frac{\kappa}{2} \ . 
\ee
 Thus, taking into account $\zeta = e^{- i \kappa}$,
\be \label{psizeta}
\Psi (x,y,\zeta) = \mm \cos\theta & i \sqrt{\zeta} \sin\theta \\  i \sin\theta /\sqrt{\zeta}   & \cos\theta \ema  C_0 \ .
\ee
Finally, using \rf{ne3}, we get the following background solution:
\be
n = \e_1 \sin 2\theta \sin \frac{\kappa}{2} - \e_2 \sin 2\theta \cos \frac{\kappa}{2} + \e_3 \cos 2 \theta \ .
\ee
Now, we will perform the Darboux-B\"acklund transformation.
In order to compute $\xi$ we evaluate $\Psi$ at $\zeta = 1/\bar \zeta_1$ and denote $\lambda_1 := 1/\sqrt{\bar \zeta_1}$:
\be
\Psi (x,y,{\bar \zeta}_1^{-1}) = \mm \cos \theta_1 & - i \lambda_1  \sin\theta_1  \\ - i \lambda_1^{-1} \sin\theta_1 & \cos\theta_1  \ema  C_0 \ ,
\ee
where $\theta_1 = \theta (x,y,\zeta_1^{-1}) \equiv P_1 + i Q_1$, i.e.,
\be  \label{P1Q1} \ba{l}  \displaystyle
P_1 = - \frac{1}{2} a_0 \left( 1 + \frac{1}{a_1^2 + b_1^2} \right)   \left( x b_1  +  y a_1 \right) \ , \\[3ex]  \displaystyle
Q_1 = \frac{1}{2} a_0 \left( 1 - \frac{1}{a_1^2 + b_1^2} \right)   \left( x a_1 - y b_1 \right) \ , 
\ea \ee
where $a_1 + i b_1 := \lambda_1 \equiv \zeta_1^{-1/2}$ (we recall that by  assumption $a_1^2 + b_1^2 \neq 1$).  
Then 
\be
\xi = \frac{c_1 - i c_2 \lambda_1 \tan\theta_1}{c_2  - i c_1 \lambda_1^{-1} \tan\theta_1} \ ,
\ee
where $(c_1,c_2)^T = C_0 \vec c $. Without loss of the generality we can put $c_1 = 0$ (one can show that more general choice is equivalent to the translation in the space of variabes $x, y$, compare \cite{Bl}). Then, finally,
\be \label{ksi}
\xi = \frac{(a_1 + i b_1)( \sinh Q_1 \cosh Q_1 - i \sin P_1 \cos P_1  )}{\cosh^2 Q_1 \cos^2 P_1 + \sinh^2 Q_1 \sin^2 P_1} \ ,
\ee
where $P_1, Q_1$ are given by \rf{P1Q1} and $a_1, b_1$ are arbitrary real parameters.

Therefore the solution $\mathbf {\tilde n}$ given by \rf{tilden} can be easily computed using \rf{DeD}, \rf{peksi}, \rf{psizeta}, \rf{ksi} and \rf{P1Q1}, where $\beta$, $a_0$, $a_1$, $b_1$, $\kappa$ and the matrix $C_0$ are arbitrary constants. In particular, assuming $C_0 = I$ and $\kappa = 0$ we obtain $\mathbf {\tilde n} = (n_1, n_2, n_3)$, where
\be  \label{n123}  \ba{l}
n_1 = 2 p_1 p_3 \sin 2\beta + p_2 \sin 2\beta \ , \\[2ex]
n_2 = (2 p_2 p_3 \sin^2 \beta - p_1 \sin 2\beta) \cos 2\theta - (\cos 2\beta + 2 p_3^2 \sin^2 \beta) \sin 2\theta \ , \\[2ex]
n_3 = (2 p_2 p_3 \sin^2 \beta - p_1 \sin 2\beta) \sin 2\theta + (\cos 2\beta + 2 p_3^2 \sin^2 \beta) \cos 2\theta \ .
\ea \ee
The functions $p_1, p_2, p_3$ are given by \rf{peksi} and \rf{ksi}, $\theta$ is given by 
\rf{teta}.

\section{Conclusions}

In this paper we presented a new version of the Darboux-B\"acklund transformation for the sigma model \rf{Nmodel}. There are two  interesting points in our construction. First, we introduced the spectral problem \rf{problin-genH}, more general than \rf{problinH}. Both spectral problems are gauge-equivalent and the sigma model \rf{Nmodel} is invariant with respect to unitary gauge transformations of the spectral problem (compare Proposition~\ref{G}). Second, the normalization matrix \rf{normalization} is quite non-trivial. The matrix ${\cal N}$ depends on $x, y$ through the projector matrix $P$ (i.e., through the background wave function). 
Note that the Darboux-B\"acklund transformation for the spectral problem \rf{problinH} is even more difficult. We have an additional constraint on the unitary matrix ${\cal N}_0$, namely \rf{additional}, which is technically pretty complicated. From this point of view the spectral problem \rf{problin-genH} is better.

Our approach is rather straightforward and we plan to generalize it for some related sigma models and geometric problems (surfaces of constant mean curvature in Euclidean and Lorentzian spaces). We hope also to extend this approach on higher dimensional problems using Clifford numbers (compare \cite{Ci-Spin}).

\end{document}